\documentclass[25pt]{article}
\usepackage{amsfonts,amsmath,amssymb,amsthm,epsfig,mathrsfs,subfigure}
\usepackage{setspace}

\theoremstyle{definition}

\theoremstyle{plain}

\usepackage{stmaryrd}

\usepackage{amssymb}
\usepackage{mathrsfs}
\usepackage{epsfig}
\usepackage{graphicx}
\usepackage{epstopdf}
\usepackage{amsfonts}
\usepackage{amsmath}
\usepackage{bm}
\usepackage{lscape}
\usepackage[indentafter]{titlesec}
\usepackage[colorlinks,linkcolor=red,linkcolor=blue,linktocpage]{hyperref}
\title{Constructing Riemann-Hilbert problem and multi-soliton solutions for the $N$-coupled Hirota equations in an optical fiber}
\author{Zhou-Zheng Kang$^{1,2}$, Tie-Cheng Xia$^{1}$\footnote{Corresponding author. E-mail: xiatc@shu.edu.cn.} \\
$^{1}$Department of Mathematics, Shanghai University, Shanghai 200444, China;\\
$^{2}$College of Mathematics, Inner Mongolia University for Nationalities,\\ Tongliao 028043, China\\}
\date{}
\begin{document}
\sloppy \maketitle
\begin{abstract}
This paper focuses on investigation of the $N$-coupled Hirota equations arising in an optical fiber. Starting from analyzing the spectral problem, a kind of matrix Riemann-Hilbert problem is formulated strictly on the real axis. Then based on the resulting matrix Riemann-Hilbert problem under the constraint of no reflection, multi-soliton solutions to the $N$-coupled Hirota equations are presented explicitly. \vskip
2mm\noindent\textbf{PACS}: 02.30.Jr, 02.30.Ik, 05.45.Yv\vskip
2mm\noindent\textbf{Keywords}: $N$-coupled Hirota equations; Riemann-Hilbert problem; soliton solutions
\end{abstract}
\section{Introduction}
As is well known, investigating soliton solutions to nonlinear evolution equaitons (NLEEs) is of an especially important significance in the study of various complex nonlinear phenomena arising from fluid dynamics, plasma physics, oceanography, optics, condensed matter physics and so forth. By now,
a variety of efficient methods have been available for seeking soliton solutions of NLEEs, some of which include the inverse scattering method [1,2], the Darboux transformation method [3--6], the B\"{a}cklund transformation method [7], the Riemann-Hilbert approach [8], and the Hirota's bilinear method [9--12]. In recent years, there has been an increasing interest in applying the Riemann-Hilbert approach to
explore abundant multi-soliton solutions of NLEEs, which include the coupled derivative Schr\"{o}dinger equation [13], the Kundu-Eckhaus equation [14], and others [15--20].

In this paper, we are concerned with the $N$-coupled Hirota equations [21]
\begin{equation}
{{q}_{jt}}=i\left[ \frac{1}{2}{{q}_{jxx}}+\left( \sum\limits_{r=1}^{N}{{{\left| {{q}_{r}} \right|}^{2}}} \right){{q}_{j}} \right]+\epsilon \left[ {{q}_{jxxx}}+3\left( \sum\limits_{r=1}^{N}{{{\left| {{q}_{r}} \right|}^{2}}} \right){{q}_{jx}}+3\left( \sum\limits_{r=1}^{N}{q_{r}^{*}{{q}_{rx}}} \right){{q}_{j}} \right],
\end{equation}
and $j=1,2,\cdots ,N,$ which governs the nonlinear wave propagation of simultaneous $N$ fields in an optical fiber with the effects of group velocity dispersion, self-phase modulation, higher-order dispersion, and self-steepening. Here $q_{j}$ represent the complex amplitude of the pulse envelope, the subscripts of $q_{j}$ denote the partial derivatives with respect to the
scaled spatial coordinate $x$ and time coordinate $t$ correspondingly,
while $\epsilon$ is a real constant, and the asterisk means the complex conjugate.
In [21], by drawing on the B\"{a}cklund transformation method, one-soliton solutions to Eqs. (1) were derived.
To the best of our knowledge, the construction of soliton solutions to Eqs. (1) has not been reported based on the Riemann-Hilbert approach, which is the main motivation of this paper.

The layout of this paper is as follows. Section 2 is devoted to the derivation of a matrix Riemann-Hilbert problem for Eqs. (1) on the real axis by analysis of the given spectral problem.
In Section 3, we will determine multi-soliton solutions to Eqs. (1) through discussing the resulting Riemann-Hilbert problem under the reflectionless case.
The final section is a short conclusion.

\section{Matrix Riemann-Hilbert problem}
In this section, we are going to put forward a matrix Riemann-Hilbert problem for Eqs. (1). We begin with the Lax pair [21]
\begin{subequations}\begin{align}
 & {{\Phi }_{x}}={{U}_{2}}\Phi =(i\lambda \Lambda +Q)\Phi , \\
 & {{\Phi }_{t}}={{V}_{2}}\Phi =\left[\left(-4i\epsilon {{\lambda }^{3}}+i{{\lambda }^{2}}\right)\Lambda +\tilde{V} \right]\Phi ,
\end{align}
\end{subequations}
where $\Phi ={{({{\Phi }_{1}},{{\Phi }_{2}},\cdots ,{{\Phi }_{N+1}})^\textrm{T}}}$ is the spectral function, $\lambda\in \mathbb{C}$ is a spectral parameter, the symbol $\textrm{T}$ means transpose of the vector, and $\tilde{V}=(-4\epsilon {{\lambda }^{2}}+\lambda )Q+\left(-2i\epsilon \lambda +\frac{i}{2}\right){{Q}_{1}}+\epsilon {{Q}_{2}},\Lambda =\text{diag}(-1,1,\cdots ,1).$
Besides,
\begin{equation*}
Q=\left( \begin{matrix}
   0 & {{q}_{1}} & {{q}_{2}} & \cdots  & {{q}_{N}}  \\
   -q_{1}^{*} & 0 & 0 & \cdots  & 0  \\
   -q_{2}^{*} & 0 & 0 & \cdots  & 0  \\
   \vdots  & \vdots  & \vdots  & \ddots  & \vdots   \\
   -q_{N}^{*} & 0 & 0 & \cdots  & 0  \\
\end{matrix} \right),\quad {{Q}_{1}}=\left( \begin{matrix}
   \sum\limits_{r=1}^{N}{{{\left| {{q}_{r}} \right|}^{2}}} & {{q}_{1x}} & {{q}_{2x}} & \cdots  & {{q}_{Nx}}  \\
   q_{1x}^{*} & -{{\left| {{q}_{1}} \right|}^{2}} & -{{q}_{2}}q_{1}^{*} & \cdots  & -{{q}_{N}}q_{1}^{*}  \\
   q_{2x}^{*} & -{{q}_{1}}q_{2}^{*} & -{{\left| {{q}_{2}} \right|}^{2}} & \cdots  & -{{q}_{N}}q_{2}^{*}  \\
   \vdots  & \vdots  & \vdots  & \ddots  & \vdots   \\
   q_{Nx}^{*} & -{{q}_{1}}q_{N}^{*} & -{{q}_{2}}q_{N}^{*} & \cdots  & -{{\left| {{q}_{N}} \right|}^{2}}  \\
\end{matrix} \right),
\end{equation*}
\[{{Q}_{2}}=\left( \begin{matrix}
   \sum\limits_{r=1}^{N}{({{q}_{rx}}q_{r}^{*}-{{q}_{r}}q_{rx}^{*})} & {{q}_{1xx}}+2{{q}_{1}}\sum\limits_{r=1}^{N}{{{\left| {{q}_{r}} \right|}^{2}}} & {{q}_{2xx}}+2{{q}_{2}}\sum\limits_{r=1}^{N}{{{\left| {{q}_{r}} \right|}^{2}}} & \cdots  & {{q}_{Nxx}}+2{{q}_{N}}\sum\limits_{r=1}^{N}{{{\left| {{q}_{r}} \right|}^{2}}} \\
   -q_{1xx}^{*}-2Aq_{1}^{*} & -({{q}_{1x}}q_{1}^{*}-{{q}_{1}}q_{1x}^{*}) & -({{q}_{2x}}q_{1}^{*}-{{q}_{2}}q_{1x}^{*}) & \cdots  & -({{q}_{Nx}}q_{1}^{*}-{{q}_{N}}q_{1x}^{*})  \\
   -q_{2xx}^{*}-2Aq_{2}^{*} & -({{q}_{1x}}q_{2}^{*}-{{q}_{1}}q_{2x}^{*}) & -({{q}_{2x}}q_{2}^{*}-{{q}_{2}}q_{2x}^{*}) & \cdots  & -({{q}_{Nx}}q_{2}^{*}-{{q}_{N}}q_{2x}^{*})  \\
   \vdots  & \vdots  & \vdots  & \ddots  & \vdots   \\
   -q_{Nxx}^{*}-2Aq_{N}^{*} & -({{q}_{1x}}q_{N}^{*}-{{q}_{1}}q_{Nx}^{*}) & -({{q}_{2x}}q_{N}^{*}-{{q}_{2}}q_{Nx}^{*}) & \cdots  & -({{q}_{Nx}}q_{N}^{*}-{{q}_{N}}q_{Nx}^{*})  \\
\end{matrix} \right).\]

Assuming that the potential functions $q_{j}$ in the Lax pair (2) decay to zero sufficiently fast as $x\rightarrow\pm\infty$, we introduce the variable transformation
\begin{equation*}
\Phi =\eta {{\text{e}}^{i\lambda \Lambda x+\left(-4i\epsilon {{\lambda }^{3}}+i{{\lambda }^{2}}\right)\Lambda t}},
\end{equation*}
under which the Lax pair (2) is turned into the desired form
\begin{subequations}
\begin{align}
 & {{\eta }_{x}}=i\lambda [\Lambda ,\eta ]+Q\eta , \\
 & {{\eta }_{t}}=\left(-4i\epsilon {{\lambda }^{3}}+i{{\lambda }^{2}}\right)[\Lambda ,\eta ]+\tilde{V}\eta ,
\end{align}
\end{subequations}
where $[\Lambda ,\eta]\equiv\Lambda \eta-\eta\Lambda$ denotes the commutator.

Now we analyze the spectral problem (3a).
Since the analysis will take place at a fixed time, the $t$-dependence will be suppressed.
With regard to (3a), we introduce its two matrix Jost solutions $\eta_{\pm}$ written as a collection of columns
\begin{equation}
{{\eta }_{-}}=({{[{{\eta }_{-}}]_{1}}},{{[{{\eta }_{-}}]_{2}}},\cdots ,{{[{{\eta }_{-}}]_{N+1}}}),\quad {{\eta }_{+}}=({{[{{\eta }_{+}}]_{1}}},{{[{{\eta }_{+}}]_{2}}},\cdots ,{{[{{\eta }_{+}}]_{N+1}}}),
\end{equation}
satisfying the large-$x$ asymptotics
\begin{equation}
{\eta_{\pm}}\to \mathbb{I}_{N+1},\quad x\to \pm\infty ,
\end{equation}
where the subscripts of $\eta$ indicated refer to which end of the $x$-axis the boundary conditions are set, and $\mathbb{I}_{N+1}$ stands for the identity matrix of rank $N+1$. The solutions ${\eta_{\pm }}$ are uniquely determined by the Volterra integral equations
\begin{subequations}
\begin{align}
 & {{\eta }_{-}}(x,\lambda)=\mathbb{I}_{N+1}+\int_{-\infty }^{x}{{{\text{e}}^{i\lambda \Lambda (x-y)}}Q(y){{\eta }_{-}}(y,\lambda){{\text{e}}^{-i\lambda \Lambda (x-y)}}\text{d}y}, \\
 & {{\eta }_{+}}(x,\lambda)=\mathbb{I}_{N+1}-\int_{x}^{+\infty }{{{\text{e}}^{i\lambda \Lambda (x-y)}}Q(y){{\eta }_{+}}(y,\lambda){{\text{e}}^{-i\lambda \Lambda (x-y)}}\text{d}y},
\end{align}
\end{subequations}
After the analysis on (6), we see that ${{[{{\eta }_{+}}]_{1}}},{{[{{\eta }_{-}}]_{2}}},\cdots ,{{[{{\eta }_{-}}]_{N+1}}}$ are analytic for $\lambda \in {\mathbb{C}^{-}}$ and continuous for $\lambda \in {\mathbb{C}^{-}}\cup \mathbb{R}$, whereas ${{[{{\eta }_{-}}]_{1}}},{{[{{\eta }_{+}}]_{2}}},\cdots ,{{[{{\eta }_{+}}]_{N+1}}}$ are analytic for $\lambda\in {\mathbb{C}^{+}}$ and continuous for $\lambda \in {\mathbb{C}^{+}}\cup \mathbb{R}$, where ${\mathbb{C}^{-}}$ and ${\mathbb{C}^{+}}$ are the lower and upper half $\lambda$-plane respectively.

Owing to $\text{tr}Q=0$, applying the Abel's identity as well as recalling the asymptotics (5), we find that $\det {\eta_{\pm }}=1$ for all $x$ and $\lambda \in \mathbb{R}.$
In addition, ${\eta_{-}}E$ and ${\eta_{+}}E$ are both the matrix solutions of the original spectral problem (2a),
they must be linearly associated, namely,
\begin{equation}
{{\eta }_{-}}E={{\eta }_{+}}ES(\lambda ),\quad E={{\text{e}}^{i\lambda \Lambda x}},
\end{equation}
where $S(\lambda)={{({{s}_{kj}})_{(N+1)\times (N+1)}}}$ is a scattering matrix, and
$\det S(\lambda)=1.$

In what follows, we shall determine two matrix functions, which are analytically continued to the upper and lower half-planes respectively. In view of the analytic properties of ${\eta_{\pm}}$, we define the first analytic function of $\lambda$ in ${\mathbb{C}^{+}}$ as
\begin{equation}
{{P}_{1}}=({{[{{\eta }_{-}}]_{1}}},{{[{{\eta }_{+}}]_{2}}},\cdots ,{{[{{\eta }_{+}}]_{N+1}}}).
\end{equation}
And then, we can obtain the asymptotic behavior
${{P}_{1}}\to \mathbb{I}_{N+1}$ as $ \lambda \in {\mathbb{C}^{+}}\to \infty .$

To establish a matrix Riemann-Hilbert problem, the analytic counterpart of $P_{1}$ in ${\mathbb{C}^{-}}$ is needed to be constructed.
we consider the adjoint scattering equation of spectral problem (3a)
\begin{equation}
{{K}_{x}}=i\lambda [\Lambda ,K]-KQ.
\end{equation}
And it is easy to know that $\eta_{\pm}^{-1}$ solve the adjoint equation (9) and obey the boundary conditions $\eta_{\pm}^{-1}\rightarrow\mathbb{I}$ as $x\rightarrow\pm\infty$. Set ${{[\eta _{\pm }^{-1}]^{l}}}$ be the $l$th row of $\eta _{\pm }^{-1}$, thus
\begin{equation}
\eta _{\pm }^{-1}=\left( \begin{matrix}
   {{[\eta _{\pm }^{-1}]^{1}}}  \\
   {{[\eta _{\pm }^{-1}]^{2}}}  \\
   \vdots   \\
   {{[\eta _{\pm }^{-1}]^{N+1}}}  \\
\end{matrix} \right).
\end{equation}
Utilizing the same techniques as before, we have
\begin{equation}
{{P}_{2}}=\left( \begin{matrix}
   {{[\eta _{-}^{-1}]^{1}}}  \\
   {{[\eta _{+}^{-1}]^{2}}}  \\
   \vdots   \\
   {{[\eta _{+}^{-1}]^{N+1}}}  \\
\end{matrix} \right),
\end{equation}
which is analytic in ${\mathbb{C}^{-}}$. Analogous to ${{P}_{1}}$, the large-$\lambda$ asymptotic behavior of ${{P}_{2}}$ turns out to be
$
{{P}_{2}}\to \mathbb{I}$ as $\lambda \in {\mathbb{C}^{-}}\to \infty .
$
From Eq. (7), we get
\begin{equation}
{{E}^{-1}}\eta _{-}^{-1}=R(\lambda ){{E}^{-1}}\eta _{+}^{-1},
\end{equation}
with $R(\lambda)={{({{r}_{kj}})_{(N+1)\times (N+1)}}}$ as the inverse matrix of $S(\lambda)$.

Insertion of (4) into Eq. (7) gives rise to
\begin{equation*}
{{[{{\eta }_{-}}]_{1}}}={{s}_{11}}{{[{{\eta }_{+}}]_{1}}}+{{s}_{21}}{{\text{e}}^{2i\lambda x}}{{[{{\eta }_{+}}]_{2}}}+{{s}_{31}}{{\text{e}}^{2i\lambda x}}{{[{{\eta }_{+}}]_{3}}}+\cdots +{{s}_{N+1,1}}{{\text{e}}^{2i\lambda x}}{{[{{\eta }_{+}}]_{N+1}}}.
\end{equation*}
Hence, ${{P}_{1}}$ takes the form
\begin{equation*}
{{P}_{1}}=({{[{{\eta }_{-}}]_{1}}},{{[{{\eta }_{+}}]_{2}}},{{[{{\eta }_{+}}]_{3}}},\cdots ,{{[{{\eta }_{+}}]_{N+1}}})=({{[{{\eta }_{+}}]_{1}}},{{[{{\eta }_{+}}]_{2}}},{{[{{\eta }_{+}}]_{3}}},\cdots ,{{[{{\eta }_{+}}]_{N+1}}})\left( \begin{matrix}
   {{s}_{11}} & 0 & 0 & \cdots  & 0  \\
   {{s}_{21}}{{\text{e}}^{2i\lambda x}} & 1 & 0 & \cdots  & 0  \\
   {{s}_{31}}{{\text{e}}^{2i\lambda x}} & 0 & 1 & \ddots  & \vdots   \\
   \vdots  & \vdots  & \ddots  & \ddots  & 0  \\
   {{s}_{N+1,1}}{{\text{e}}^{2i\lambda x}} & 0 & \cdots  & 0 & 1  \\
\end{matrix} \right),
\end{equation*}

Through substituting (10) into Eq. (12), we derive
\begin{equation*}
{{[\eta _{-}^{-1}]^{1}}}={{r}_{11}}{{[\eta _{+}^{-1}]^{1}}}+{{r}_{12}}{{\text{e}}^{-2i\lambda x}}{{[\eta _{+}^{-1}]^{2}}}+{{r}_{13}}{{\text{e}}^{-2i\lambda x}}{{[\eta _{+}^{-1}]^{3}}}+\cdots +{{r}_{1,N+1}}{{\text{e}}^{-2i\lambda x}}{{[\eta _{+}^{-1}]^{N+1}}}.
\end{equation*}
Subsequently, ${{P}_{2}}$ is represented as
\begin{equation*}
{{P}_{2}}=\left( \begin{matrix}
   {{[\eta _{-}^{-1}]^{1}}}  \\
   {{[\eta _{+}^{-1}]^{2}}}  \\
   {{[\eta _{+}^{-1}]^{3}}}  \\
   \vdots   \\
   {{[\eta _{+}^{-1}]}^{N+1}}  \\
\end{matrix} \right)=\left( \begin{matrix}
   {{r}_{11}} & {{r}_{12}}{{\text{e}}^{-2i\lambda x}} & {{r}_{13}}{{\text{e}}^{-2i\lambda x}} & \cdots  & {{r}_{1,N+1}}{{\text{e}}^{-2i\lambda x}}  \\
   0 & 1 & 0 & \cdots  & 0  \\
   0 & 0 & 1 & \ddots  & \vdots   \\
   \vdots  & \vdots  & \ddots  & \ddots  & 0  \\
   0 & 0 & \cdots  & 0 & 1  \\
\end{matrix} \right)\left( \begin{matrix}
   {{[\eta _{+}^{-1}]^{1}}}  \\
   {{[\eta _{+}^{-1}]^{2}}}  \\
   {{[\eta _{+}^{-1}]^{3}}}  \\
   \vdots   \\
   {{[\eta _{+}^{-1}]^{N+1}}}  \\
\end{matrix} \right),
\end{equation*}

Having constructed two matrix functions ${{P}_{1}}$ and ${{P}_{2}}$ which are analytic in ${\mathbb{C}^{+}}$ and ${\mathbb{C}^{-}}$ respectively,
we are ready to describe a Riemann-Hilbert problem for Eqs. (1).
After denoting that the limit of ${{P}_{1}}$ is ${{P}^{+}}$ as $\lambda\in {\mathbb{C}^{+}}\rightarrow\mathbb{R}$ and
the limit of ${{P}_{2}}$ is ${{P}^{-}}$ as $\lambda \in {\mathbb{C}^{-}}\rightarrow\mathbb{R}$, a matrix Riemann-Hilbert problem desired can be presented below
\begin{equation}
{{P}^{-}}(x,\lambda){{P}^{+}}(x,\lambda)=G(x,\lambda),\quad \lambda \in \mathbb{R},
\end{equation}
and
\begin{equation*}
G(x,\lambda)=\left( \begin{matrix}
   1 & {{r}_{12}}{{\text{e}}^{-2i\lambda x}} & {{r}_{13}}{{\text{e}}^{-2i\lambda x}} & \cdots  & {{r}_{1,N+1}}{{\text{e}}^{-2i\lambda x}}  \\
   {{s}_{21}}{{\text{e}}^{2i\lambda x}} & 1 & 0 & \cdots  & 0  \\
   {{s}_{31}}{{\text{e}}^{2i\lambda x}} & 0 & 1 & \ddots  & \vdots   \\
   \vdots  & \vdots  & \ddots  & \ddots  & 0  \\
   {{s}_{N+1,1}}{{\text{e}}^{2i\lambda x}} & 0 & \cdots  & 0 & 1  \\
\end{matrix} \right),
\end{equation*}
with its canonical normalization conditions given by
\begin{eqnarray*}
  &{{P}_{1}}(x,\lambda)\to \mathbb{I}_{N+1},\quad \lambda \in {\mathbb{C}^{+}}\to \infty , \\
  &{{P}_{2}}(x,\lambda)\to \mathbb{I}_{N+1},\quad \lambda \in {\mathbb{C}^{-}}\to \infty ,
\end{eqnarray*}
and ${{r}_{11}}{{s}_{11}}+{{r}_{12}}{{s}_{21}}+\cdots +{{r}_{1,N+1}}{{s}_{N+1,1}}=1$.

\section{Soliton solutions}
In this section, we construct multi-soliton solutions to Eqs. (1) based on the Riemann-Hilbert problem presented above.
We now suppose the Riemann-Hilbert problem (13) to be irregular, which signifies that both $\det {{P}_{1}}$ and $\det {{P}_{2}}$ are in possession of some zeros in analytic domains of their own.
According to the definitions of ${{P}_{1}}$ and ${{P}_{2}}$ as well as the scattering relation between ${{\eta}_{-}}$ and ${{\eta}_{+}}$ in Eq. (7), we find that
\begin{equation*}
\det {{P}_{1}}(\lambda)={{s}_{11}}(\lambda),\quad
\det {{P}_{2}}(\lambda)={{r}_{11}}(\lambda),
\end{equation*}
which show us that
$\det {{P}_{1}}$ and $\det {{P}_{2}}$ possess the same zeros as ${s}_{11}$ and ${r}_{11}$ respectively.

With above analysis, we now reveal the characteristic feature of zeros.
It is noted that the potential matrix $Q$ is anti-Hermitian, namely,
$
Q^{\dagger }=-Q.
$
On basis of this relation, we deduce that
\begin{equation}
\eta_{\pm }^{\dagger }({{\lambda}^{*}})=\eta_{\pm }^{-1}(\lambda).
\end{equation}
In order to ease discussion, we introduce two special matrices
\[{{H}_{1}}=\text{diag}(1,\underbrace{0,0,\cdots ,0}_{N}),\quad {{H}_{2}}=\text{diag}(0,\underbrace{1,1,\cdots ,1}_{N}),\]
which allows us to express (8) and (11) as
\begin{equation}
{{P}_{1}}={{\eta }_{-}}{{H}_{1}}+{{\eta }_{+}}{{H}_{2}},\quad {{P}_{2}}={{H}_{1}}\eta _{-}^{-1}+{{H}_{2}}\eta _{+}^{-1},
\end{equation}
Taking the Hermitian of the first equation of (15) and making use of the relation (14), we find that
\begin{equation}
P_{1}^{\dagger }({{\lambda}^{*}})={{P}_{2}}(\lambda),\quad {{S}^{\dagger }}({{\lambda}^{*}})={{S}^{-1}}(\lambda),
\end{equation}
for $\lambda \in {\mathbb{C}^{-}}.$
From the second equation of (16), we further have
$s_{11}^{*}({{\lambda}^{*}})={{r}_{11}}(\lambda)$.
Therefore, we assume that
$\det {{P}_{1}}$ has $n$ simple zeros $\{{{\lambda}_{l}}\}_{l=1}^{n}$ in ${\mathbb{C}^{+}}$
and $\det {{P}_{2}}$ has $n$ simple zeros $\{{{\hat{\lambda}}_{l}}\}_{l=1}^{n}$ in ${\mathbb{C}^{-}}$, where
${{\hat{\lambda}}_{l}}=\lambda_{l}^{*}.$
Each of $\ker {{P}_{1}}({{\lambda }_{l}})$ includes only a single basis column vector ${{\nu}_{l}}$, and each of $\ker {{P}_{2}}({{\hat{\lambda }}_{l}})$ includes only a single basis row vector ${{\hat{\nu}}_{l}}$,
\begin{equation}
{{P}_{1}}({{\lambda}_{l}}){{\nu}_{l}}=0,\quad {{\hat{\nu}}_{l}}{{P}_{2}}({{\hat{\lambda}}_{l}})=0.
\end{equation}
Taking the Hermitian of the first equation of (17) and using (16), we find that the eigenvectors fulfill the relation
\begin{equation}
{{\hat{\nu}}_{l}}=\nu_{l}^{\dagger },\quad 1\le l\le n.
\end{equation}
Differentiating the first equation of (17) in $x$ and $t$ respectively and taking advantage of the Lax pair (3), we arrive at
\begin{equation*}
{{P}_{1}}({{\lambda }_{l}})\left( \frac{\partial {{\nu}_{l}}}{\partial x}-i{{\lambda }_{l}}\Lambda {{\nu}_{l}} \right)=0,\quad
{{P}_{1}}({{\lambda }_{l}})\left( \frac{\partial {{\nu}_{l}}}{\partial t}-\left( i\lambda _{l}^{2}-4i\epsilon \lambda _{l}^{3} \right)\Lambda {{\nu }_{l}} \right)=0,
\end{equation*}
which leads to
\begin{equation}
{{\nu}_{l}}={{\text{e}}^{i{{\lambda }_{l}}\Lambda x+\left(i\lambda _{l}^{2}-4i\epsilon \lambda _{l}^{3}\right)\Lambda t}}{{\nu}_{l,0}},\quad 1\le l\le n,
\end{equation}
with ${{\nu}_{l,0}}$ being independent of the variables $x$ and $t$. In consideration of the relation (18), we thus have
\begin{equation}
{{\hat{\nu}}_{l}}=\nu_{l,0}^{\dagger }{{\text{e}}^{-i\lambda _{l}^{*}\Lambda x+\left(4i\epsilon \lambda {{_{l}^{*}}^{3}}-i\lambda {{_{l}^{*}}^{2}}\right)\Lambda t}},\quad 1\le l\le n.
\end{equation}

In order to derive soliton solutions explicitly, we here take $G=\mathbb{I}_{N+1}$, which means that no reflection exists in the scattering problem. Therefore, the solutions [22] to the Riemann-Hilbert problem (13) can be given by
\begin{equation}
{{P}_{1}}(\lambda)=\mathbb{I}_{N+1}-\sum\limits_{k=1}^{n}{\sum\limits_{l=1}^{n}{\frac{{{\nu}_{k}}{{{\hat{\nu}}}_{l}}{{\big({{M}^{-1}}\big)}_{kl}}}{\lambda -{{{\hat{\lambda}}}_{l}}}}},\quad
{{P}_{2}}(\lambda)=\mathbb{I}_{N+1}+\sum\limits_{k=1}^{n}{\sum\limits_{l=1}^{n}{\frac{{{\nu}_{k}}{{{\hat{\nu}}}_{l}}{{\big({{M}^{-1}}\big)}_{kl}}}{\lambda-{{\lambda }_{k}}}}},
\end{equation}
in which $M$ is defined as
\begin{equation}
M=({{M}_{kl}})_{n\times n}=\left(\frac{{\hat{\nu}_{k}}{{{{\nu}}}_{l}}}{{{\lambda}_{l}}-{{{\hat{\lambda}}}_{k}}}\right)_{n\times n},\quad 1\le k,l\le n,
\end{equation}
and ${{\big({{M}^{-1}}\big)}_{kl}}$ stands for the $(k,l)$-element of the inverse matrix of $M$.

Since $P_{1}$ satisfies the spectral problem (3a), we substitute
\begin{equation*}
{{P}_{1}}(\lambda)=\mathbb{I}_{N+1}+{{\lambda }^{-1}}P_{1}^{(1)}+{{\lambda }^{-2}}P_{1}^{(2)}+O\big({{\lambda }^{-3}}\big),\quad \lambda \to \infty ,
\end{equation*}
into (3a) and generate
\begin{equation*}
Q=-i\big[\Lambda ,P_{1}^{(1)}\big]=\left( \begin{matrix}
   0 & 2i{{\big(P_{1}^{(1)}\big)_{12}}} & \cdots & 2i{{\big(P_{1}^{(1)}\big)_{1,N+1}}}  \\
   -2i{{\big(P_{1}^{(1)}\big)_{21}}} & 0 & \cdots & 0  \\
   \vdots  & \vdots  & \ddots  & \vdots   \\
   -2i{{\big(P_{1}^{(1)}\big)_{N+1,1}}} & 0 & \cdots & 0  \\
\end{matrix} \right).
\end{equation*}
Hence the potential functions are constructed as
\begin{equation}
{{q}_{j}}=2i{{\big(P_{1}^{(1)}\big)_{1,j+1}}},\quad j=1,2,\cdots ,N.
\end{equation}
Now from Eq. (21), we obtain
\begin{equation*}
P_{1}^{(1)}=-\sum\limits_{k=1}^{n}{\sum\limits_{l=1}^{n}{{{\nu}_{k}}{{{\hat{\nu}}}_{l}}{{\big({{M}^{-1}}\big)_{kl}}}}}.
\end{equation*}
As a result, multi-soliton solutions to Eqs. (1) we are searching for can be written as
\begin{equation*}
{{q}_{j}}=-2i\sum\limits_{k=1}^{n}{\sum\limits_{l=1}^{n}{{{\nu}_{k,1}}{{{\hat{\nu}}}_{l,j+1}}{{\big({{M}^{-1}}\big)}_{kl}}}},\quad j=1,2,\cdots ,N,
\end{equation*}
in which the matrix $M$ is given by (22), and ${{\nu}_{k}}={{\left( {{\nu}_{k,1}},{{\nu}_{k,2}},\cdots ,{{\nu}_{k,N+1}} \right)^\textrm{T}}}$ and ${{\hat{\nu}}_{k}}={{\left( {{{\hat{\nu}}}_{k,1}},{{{\hat{\nu}}}_{k,2}},\cdots ,{{{\hat{\nu}}}_{k,N+1}} \right)}},1\le k\le n,$ are determined by (19) and (20).

As a particular reduction, we now take $N=3$ in Eqs. (1), which gives rise to the three-coupled Hirota equations
\begin{equation}
{{q}_{jt}}=i\left[ \frac{1}{2}{{q}_{jxx}}+\left( \sum\limits_{r=1}^{3}{{{\left| {{q}_{r}} \right|}^{2}}} \right){{q}_{j}} \right]+\epsilon \left[ {{q}_{jxxx}}+3\left( \sum\limits_{r=1}^{3}{{{\left| {{q}_{r}} \right|}^{2}}} \right){{q}_{jx}}+3\left( \sum\limits_{r=1}^{3}{q_{r}^{*}{{q}_{rx}}} \right){{q}_{j}} \right],
\end{equation}
where $j=1,2,3.$ The formulas established above can be employed to write out its explicit expressions of multi-soliton solutions. In particular, the bell-shaped one-soliton solutions are as follows
\begin{equation}
\begin{aligned}
 & {{q}_{1}}=\frac{-2i{{\alpha }_{1}}\beta _{1}^{*}{{\text{e}}^{\theta _{1}^{*}-{{\theta }_{1}}}}\left({{\lambda }_{1}}-\lambda _{1}^{*} \right)}{{{\left| {{\alpha }_{1}} \right|}^{2}}{{\text{e}}^{-\theta _{1}^{*}-{{\theta }_{1}}}}+\big( {{\left| {{\beta }_{1}} \right|}^{2}}+{{\left| {{\gamma }_{1}} \right|}^{2}}+{{\left| {{\delta }_{1}} \right|}^{2}} \big){{\text{e}}^{\theta _{1}^{*}+{{\theta }_{1}}}}}, \\
 & {{q}_{2}}=\frac{-2i{{\alpha }_{1}}\gamma _{1}^{*}{{\text{e}}^{\theta _{1}^{*}-{{\theta }_{1}}}}\left({{\lambda }_{1}}-\lambda _{1}^{*} \right)}{{{\left| {{\alpha }_{1}} \right|}^{2}}{{\text{e}}^{-\theta _{1}^{*}-{{\theta }_{1}}}}+\big( {{\left| {{\beta }_{1}} \right|}^{2}}+{{\left| {{\gamma }_{1}} \right|}^{2}}+{{\left| {{\delta }_{1}} \right|}^{2}} \big){{\text{e}}^{\theta _{1}^{*}+{{\theta }_{1}}}}}, \\
 & {{q}_{3}}=\frac{-2i{{\alpha }_{1}}\delta _{1}^{*}{{\text{e}}^{\theta _{1}^{*}-{{\theta }_{1}}}}\left({{\lambda }_{1}}-\lambda _{1}^{*} \right)}{{{\left| {{\alpha }_{1}} \right|}^{2}}{{\text{e}}^{-\theta _{1}^{*}-{{\theta }_{1}}}}+\big( {{\left| {{\beta }_{1}} \right|}^{2}}+{{\left| {{\gamma }_{1}} \right|}^{2}}+{{\left| {{\delta }_{1}} \right|}^{2}} \big){{\text{e}}^{\theta _{1}^{*}+{{\theta }_{1}}}}},
\end{aligned}
\end{equation}
where ${{\theta }_{1}}=i{{\lambda }_{1}}x+\big( i\lambda _{1}^{2}-4i\epsilon \lambda _{1}^{3}\big)t.$ This set of solutions then reduces to the more concise forms
\begin{equation}
\begin{aligned}
 & {{q}_{1}}=2\beta _{1}^{*}{{b}_{1}}{{\text{e}}^{\theta _{1}^{*}-{{\theta }_{1}}}}{{\text{e}}^{-{{\xi }_{1}}}}\text{sech(}\theta _{1}^{*}+{{\theta }_{1}}+{{\xi }_{1}}\text{),} \\
 & {{q}_{2}}=2\gamma _{1}^{*}{{b}_{1}}{{\text{e}}^{\theta _{1}^{*}-{{\theta }_{1}}}}{{\text{e}}^{-{{\xi }_{1}}}}\text{sech(}\theta _{1}^{*}+{{\theta }_{1}}+{{\xi }_{1}}\text{)}, \\
 & {{q}_{3}}=2\delta _{1}^{*}{{b}_{1}}{{\text{e}}^{\theta _{1}^{*}-{{\theta }_{1}}}}{{\text{e}}^{-{{\xi }_{1}}}}\text{sech(}\theta _{1}^{*}+{{\theta }_{1}}+{{\xi }_{1}}\text{)},
\end{aligned}
\end{equation}
via specifying the parameter ${{\alpha }_{1}}=1,$ and assuming ${{\lambda }_{1}}={a}_{1}+i{b}_{1},{{\left| {{\beta }_{1}} \right|}^{2}}+{{\left| {{\gamma }_{1}} \right|}^{2}}+{{\left| {{\delta }_{1}} \right|}^{2}}={{\text{e}}^{2{{\xi }_{1}}}}$.
From the solutions (26), it is easily found that the peak amplitudes of ${{q}_{1}},{{q}_{2}},$ and ${{q}_{3}}$ are $2{\left| \beta _{1}^{*} \right|}{{b}_{1}}{{\text{e}}^{-{{\xi }_{1}}}},2{\left| \gamma _{1}^{*} \right|}{{b}_{1}}{{\text{e}}^{-{{\xi }_{1}}}},$ and $2{\left|\delta_{1}^{*}\right|}{{b}_{1}}{{\text{e}}^{-{{\xi }_{1}}}},$ respectively.
In view of the similar structures of solutions ${{q}_{1}},{{q}_{2}}$, and ${{q}_{3}}$, we here choose ${{q}_{1}}$ as an example to display its localized structures by figures.
\begin{figure}
\begin{center}
\subfigure[]{\includegraphics[width=1.9in]{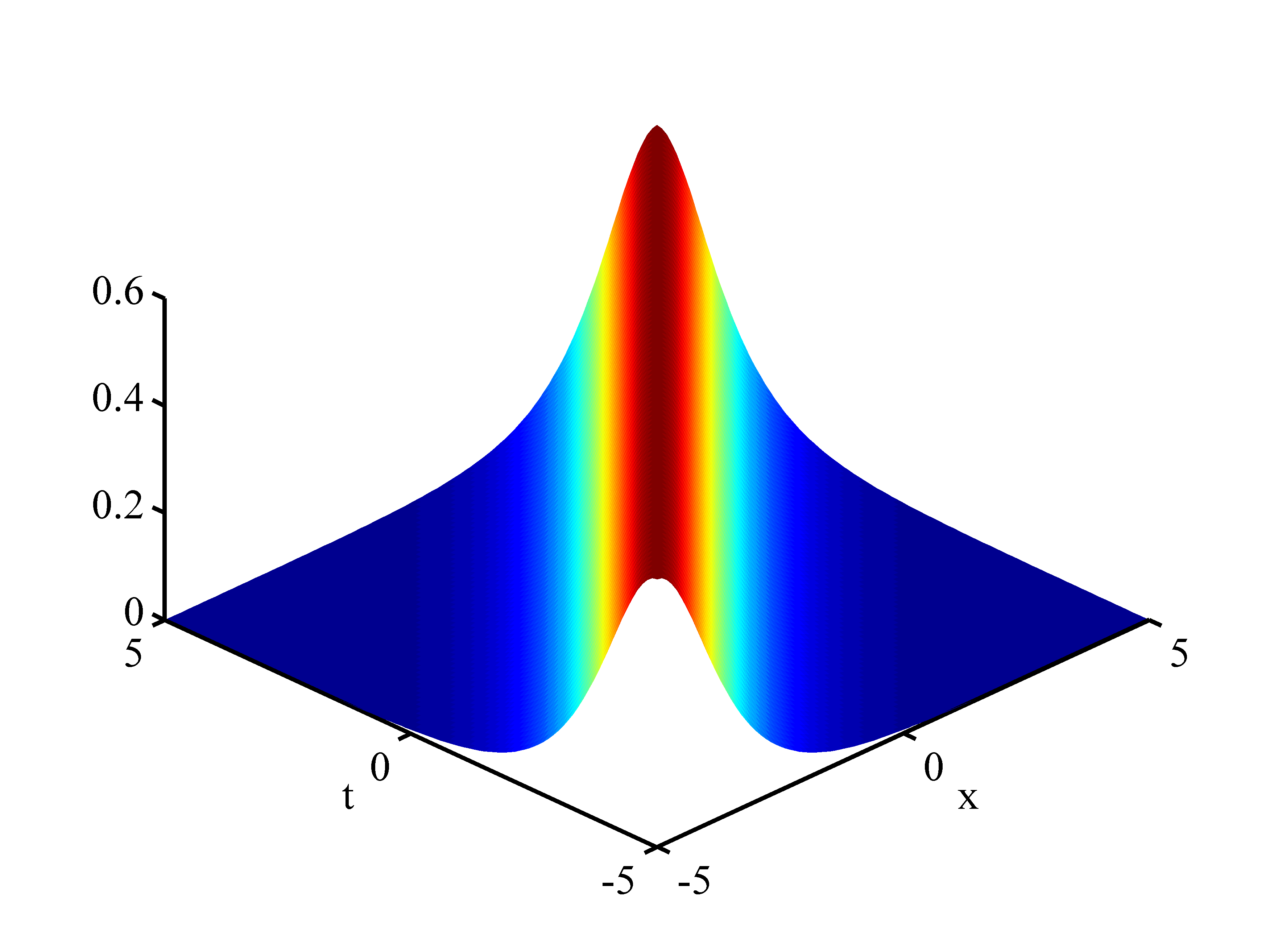}}
\subfigure[]{\includegraphics[width=1.9in]{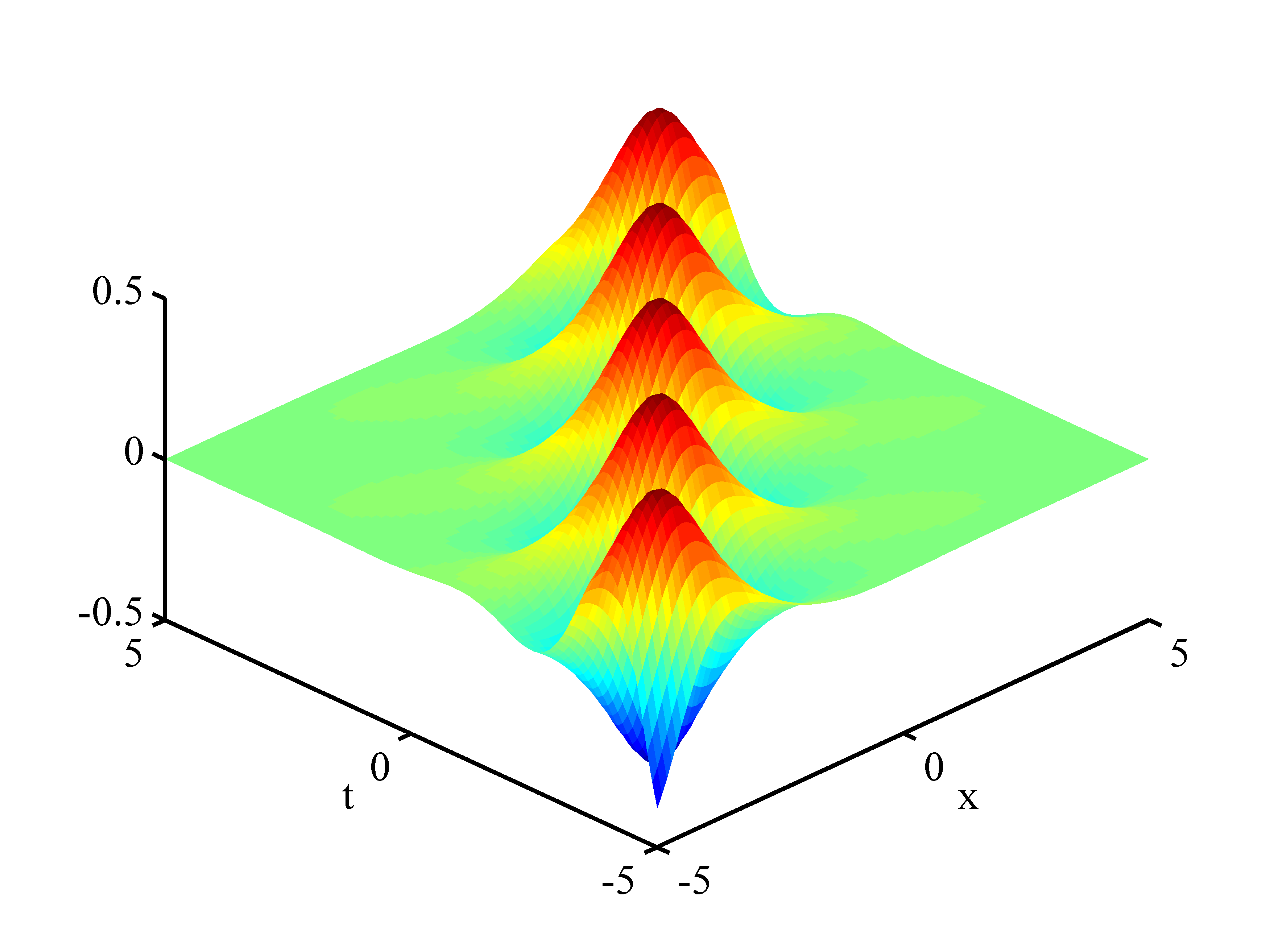}}
\subfigure[]{\includegraphics[width=1.9in]{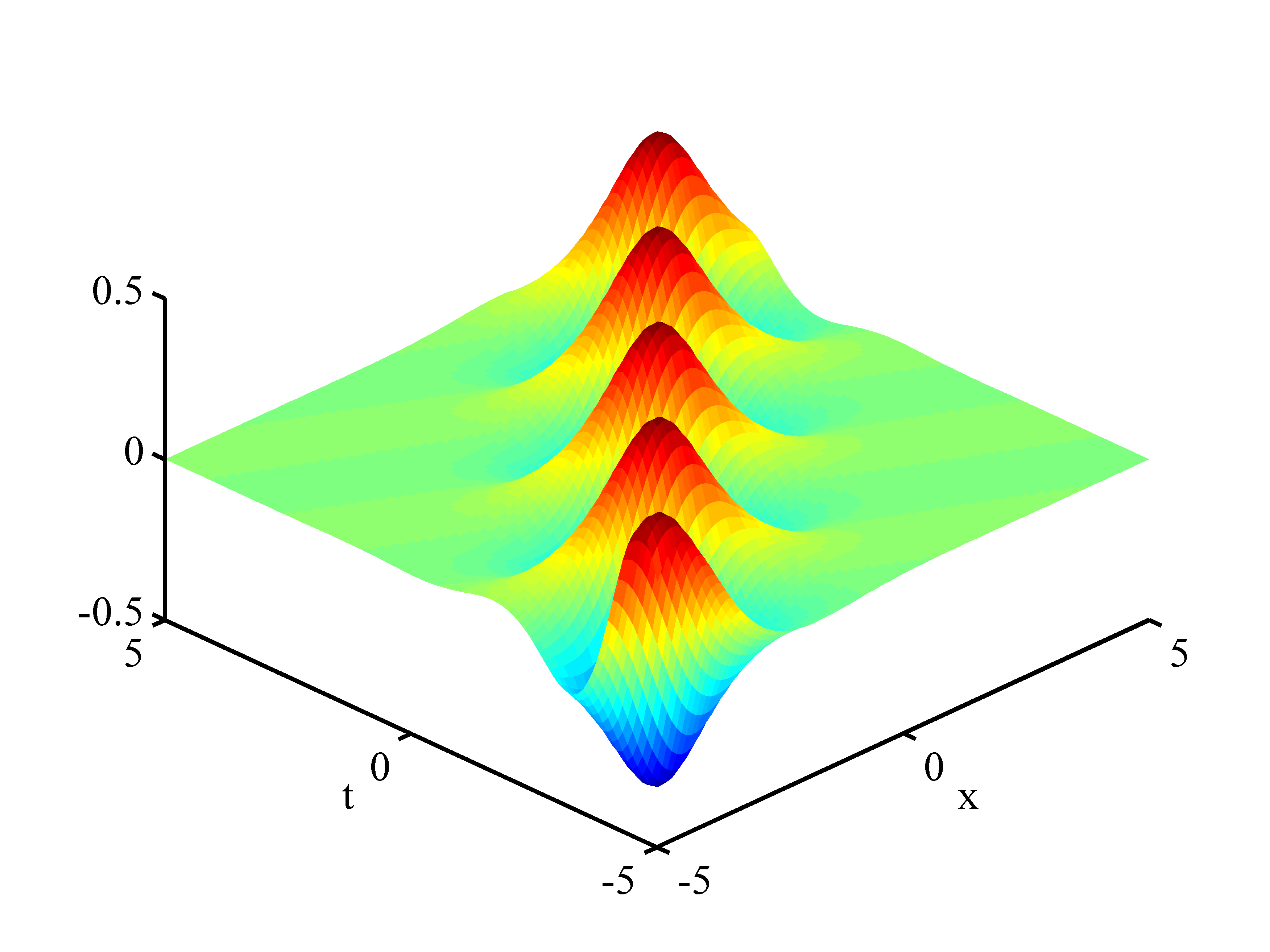}}
\parbox[c]{14cm}{\footnotesize
{\bf Figure 1.}~Plots of one-soliton solution ${q}_{1}$ in (26) with ${{\alpha }_{1}}=\epsilon=1,{{\beta }_{1}}=0.5,{{\gamma }_{1}}=0.2,{{\xi}_{1}}=0,{{\lambda }_{1}}=0.5+0.5i$. (a) perspective view of modulus of ${q}_{1}$; (b) perspective view of real part of ${q}_{1}$; (c) perspective view of imaginary part of ${q}_{1}$.}
\end{center}
\addtocounter{subfigure}{-3}
\end{figure}
\begin{figure}
\begin{center}
\subfigure[]{\includegraphics[width=1.9in]{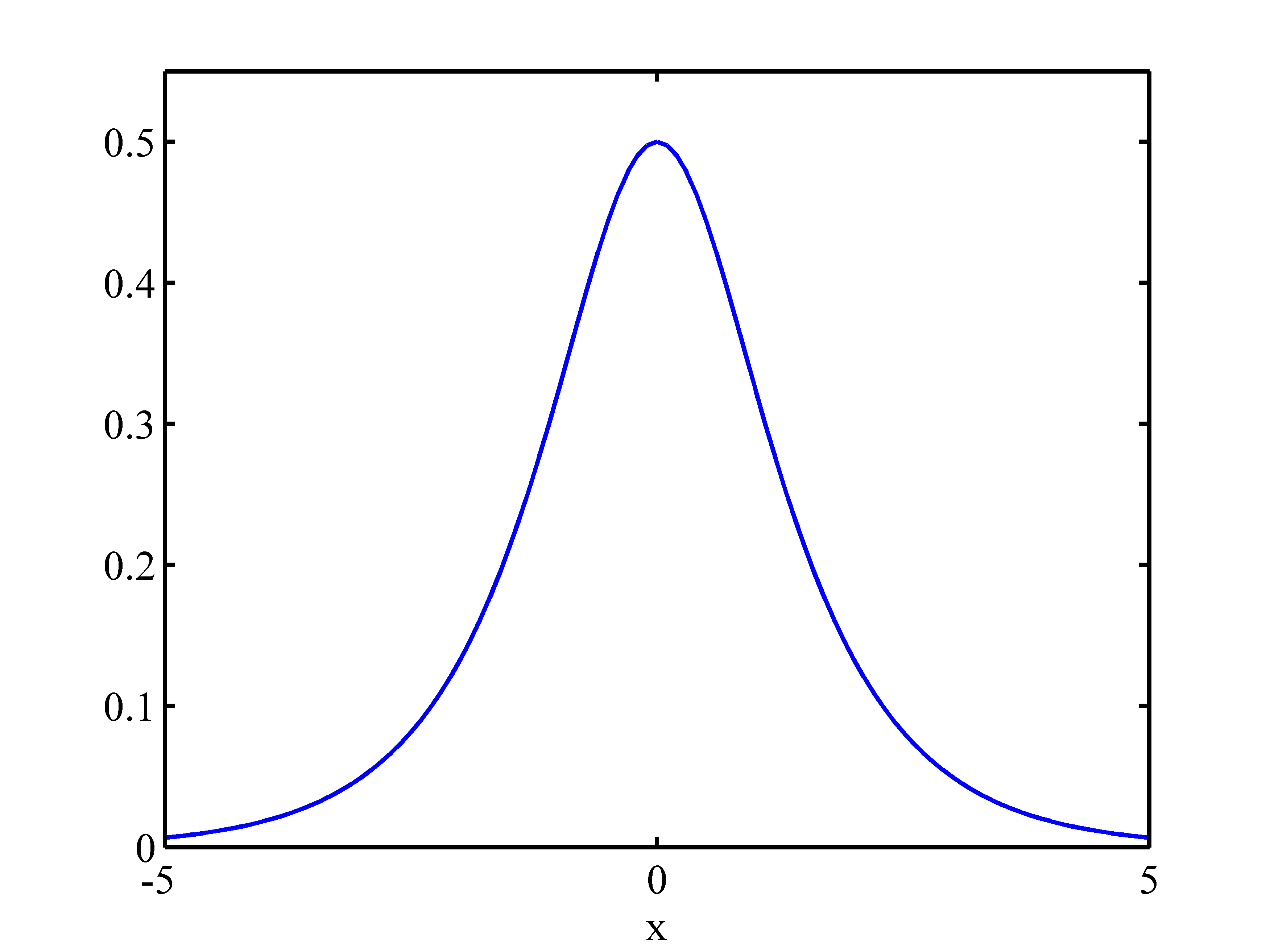}}
\subfigure[]{\includegraphics[width=1.9in]{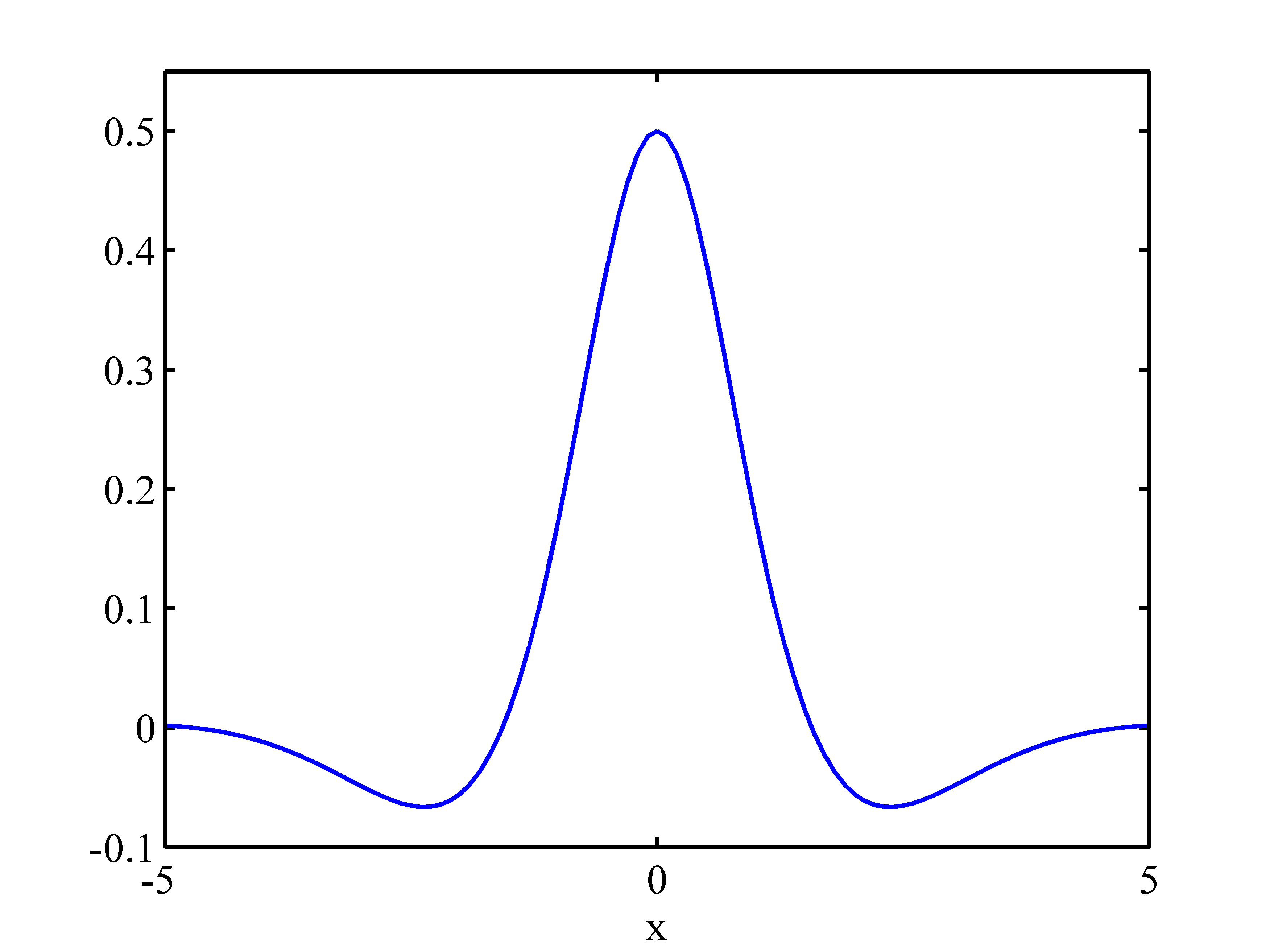}}
\subfigure[]{\includegraphics[width=1.9in]{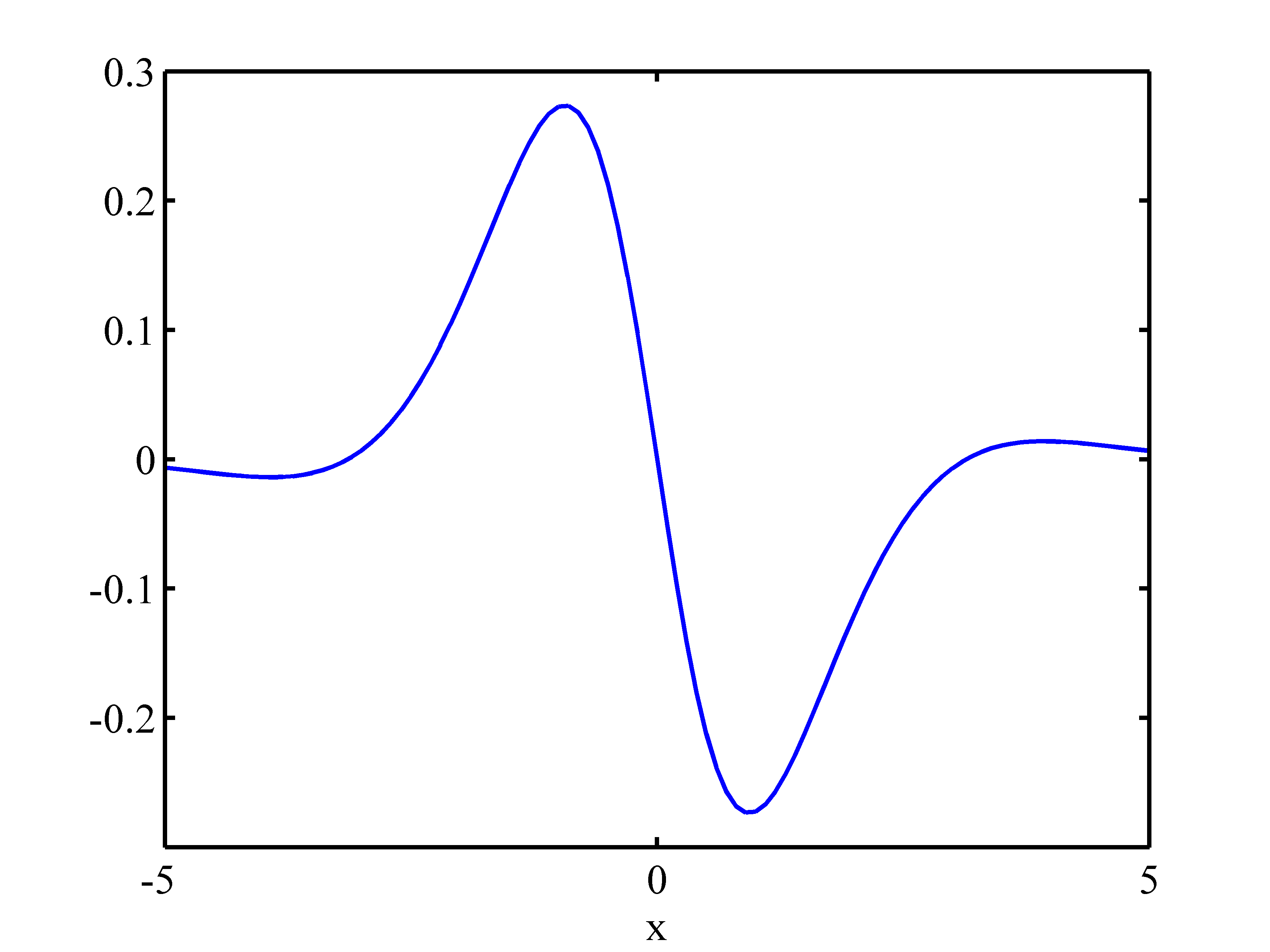}}
\parbox[c]{14cm}{\footnotesize
{\bf Figure 2.}~Plots of one-soliton solution ${q}_{1}$ in (26) with ${{\alpha }_{1}}=\epsilon=1,{{\beta }_{1}}=0.5,{{\gamma }_{1}}=0.2,{{\xi}_{1}}=0,{{\lambda }_{1}}=0.5+0.5i,t=0$. (a) $x$-curve of modulus of ${q}_{1}$; (b) $x$-curve of real part of ${q}_{1}$; (c) $x$-curve of imaginary part of ${q}_{1}$.}
\end{center}
\addtocounter{subfigure}{-3}
\end{figure}

\section{Conclusion}
The aim of the present research was to seek multi-soliton solutions for the $N$-coupled Hirota equations arising in an optical fiber. To this end, we firstly performed the spectral analysis and formulated a kind of matrix Riemann-Hilbert problem on the real axis. Secondly, based on the resulting Riemann-Hilbert problem which was treated by considering that no reflection exists in the scattering problem, the expressions of general multi-soliton solutions for the $N$-coupled Hirota equations were presented explicitly. Moreover, as a by-product, the three-coupled Hirota equations together its bell-shaped one-soliton solutions were generated. Future research should be undertaken to investigate other multiple coupled equations.

\section*{Acknowledgements}
This work was supported by the National Natural Science Foundation of China (Grant Nos. 61072147 and 11271008).



\end{document}